\documentclass[%
reprint, superscriptaddress,
 amsmath,amssymb,
 Aps, 
prd,
longbibliography,
]{revtex4-1}
\AtBeginDocument{%
    \newwrite\bibnotes
    \def\bibnotesext{Notes.bib}
    \immediate\openout\bibnotes=\jobname\bibnotesext
    \immediate\write\bibnotes{@CONTROL{REVTEX41Control}}
    \immediate\write\bibnotes{@CONTROL{%
    apsrev41Control,author="08",editor="1",pages="1",title="1",year="1"}}
     \if@filesw
     \immediate\write\@auxout{\string\citation{apsrev41Control}}%
    \fi
}%
\usepackage{graphicx}
\usepackage{dcolumn}
\usepackage{bm}
\DeclareUnicodeCharacter{0308}{HERE!HERE!}
\usepackage{hyperref}

\begin{document}

\preprint{APS/123-QED}

\title{Switching dynamics in Al/InAs nanowire-based gate-controlled superconducting switch}

\author{Tosson Elalaily}
\affiliation{Department of Physics, Institute of Physics, Budapest University of Technology and Economics, M\"uegyetem rkp. 3., H-1111 Budapest, Hungary\\}
\affiliation{MTA-BME Superconducting Nanoelectronics Momentum Research Group, M\"uegyetem rkp. 3., H-1111 Budapest, Hungary\\}
\affiliation{Department of Physics, Faculty of Science, Tanta University, Al-Geish St., 31527 Tanta, Gharbia, Egypt\\}

\author{Martin Berke}
\affiliation{Department of Physics, Institute of Physics, Budapest University of Technology and Economics, M\"uegyetem rkp. 3., H-1111 Budapest, Hungary\\}
\affiliation{MTA-BME Superconducting Nanoelectronics Momentum Research Group, M\"uegyetem rkp. 3., H-1111 Budapest, Hungary\\}

\author{Ilari Lilja}
\affiliation{Low-Temperature Laboratory, Department of Applied Physics, Aalto University School of Science, P.O. Box 15100, Fi-00076 Aalto, Finland\\}
\affiliation{QTF Centre of Excellence, Department of Applied Physics, Aalto University, Fi-00076 Aalto, Finland\\}

\author{Alexander Savin}
\affiliation{QTF Centre of Excellence, Department of Applied Physics, Aalto University, Fi-00076 Aalto, Finland\\}

\author{Gerg\H{o} F\"ul\"op}
\affiliation{Department of Physics, Institute of Physics, Budapest University of Technology and Economics, M\"uegyetem rkp. 3., H-1111 Budapest, Hungary\\}
\affiliation{MTA-BME Superconducting Nanoelectronics Momentum Research Group, M\"uegyetem rkp. 3., H-1111 Budapest, Hungary\\}

\author{L\H{o}rinc Kup\'as}
\affiliation{Department of Physics, Institute of Physics, Budapest University of Technology and Economics, M\"uegyetem rkp. 3., H-1111 Budapest, Hungary\\}
\affiliation{MTA-BME Superconducting Nanoelectronics Momentum Research Group, M\"uegyetem rkp. 3., H-1111 Budapest, Hungary\\}

\author{Thomas Kanne}
 \affiliation{Center for Quantum Devices and Nano-Science Center, Niels Bohr Institute, University of Copenhagen, Universitetsparken 5, DK-2100, Copenhagen, Denmark\\}
 
\author{Jesper Nygård}
 \affiliation{Center for Quantum Devices and Nano-Science Center, Niels Bohr Institute, University of Copenhagen, Universitetsparken 5, DK-2100, Copenhagen, Denmark\\}

\author{P\'eter Makk}
\email{Makk.peter@ttk.bme.hu}
\affiliation{Department of Physics, Institute of Physics, Budapest University of Technology and Economics, M\"uegyetem rkp. 3., H-1111 Budapest, Hungary\\}
\affiliation{MTA-BME Correlated van der Waals Structures Momentum Research Group, M\"uegyetem rkp. 3., H-1111 Budapest, Hungary\\}

\author{Pertti Hakonen}
\email{Pertti.hakonen@aalto.fi}
\affiliation{Low-Temperature Laboratory, Department of Applied Physics, Aalto University School of Science, P.O. Box 15100, Fi-00076 Aalto, Finland\\}
\affiliation{QTF Centre of Excellence, Department of Applied Physics, Aalto University, Fi-00076 Aalto, Finland\\}

\author{Szabolcs Csonka}
\affiliation{Department of Physics, Institute of Physics, Budapest University of Technology and Economics, M\"uegyetem rkp. 3., H-1111 Budapest, Hungary\\}
\affiliation{MTA-BME Superconducting Nanoelectronics Momentum Research Group, M\"uegyetem rkp. 3., H-1111 Budapest, Hungary\\}
\affiliation{Institute of Technical Physics and Materials Science, Center for Energy Research, Konkoly-Thege Mikl\'os \'ut 29-33., 1121, Budapest, Hungary\\}
\date{\today}

\begin{abstract}
The observation of the gate-controlled supercurrent (GCS) effect in superconducting nanostructures increased the hopes for realizing a superconducting equivalent of semiconductor field-effect transistors. However, recent works attribute this effect to various leakage-based scenarios, giving rise to a debate on its origin. A proper understanding of the microscopic process underlying the GCS effect and the relevant time scales would be beneficial to evaluate the possible applications. In this work, we observed gate-induced two-level fluctuations between the superconducting state and normal state in Al/InAs nanowires (NWs). Noise correlation measurements show a strong correlation with leakage current fluctuations. The time-domain measurements show that these fluctuations have Poissonian statistics. Our detailed analysis of the leakage current measurements reveals that it is consistent with the stress-induced leakage current (SILC), in which inelastic tunneling with phonon generation is the predominant transport mechanism. Our findings shed light on the microscopic origin of the GCS effect and give deeper insight into the switching dynamics of the superconducting NW under the influence of the strong gate voltage. 
\end{abstract}

\maketitle


\section{Introduction}
\label{sec:Introduction}
 In 2018, De Simoni et al.\,\cite{de2018metallic} observed a monotonic suppression of the supercurrent in titanium (Ti) NWs by increasing the voltage beyond a certain threshold at a nearby gate (side or back) electrode. With further increases in the gate voltage, the supercurrent is fully suppressed, and the device is switched to the normal state, providing a superconducting equivalent of semiconductor CMOS transistors with an expected high switching speed and low power consumption. Moreover, the structure of gate-controlled superconducting nanodevices can be readily scaled up with greater compatibility for interfacing with CMOS transistors in comparison to alternative superconducting devices \cite{buck1956cryotron,matisoo1966subnanosecond,likharev1991rsfq,mccaughan2014superconducting}. The earlier experimental observations of the GCS were associated with various unique fingerprints, which were attributed to the influence of the external $E$-field \cite{de2018metallic,de2019josephson,paolucci2019magnetotransport,rocci2020gate,puglia2020electrostatic,bours2020unveiling,mercaldo2020electrically,solinas2021sauter,mercaldo2021spectroscopic,chirolli2021impact,amoretti2022destroying,chakraborty2023microscopic}. Later on, most of these fingerprints were observed in other works suggesting leakage current-based scenarios \cite{alegria2021high,ritter2021superconducting,golokolenov2021origin,elalaily2021gate,ritter2022out,elalaily2023signature,basset2021gate}. Furthermore, the different experimental conditions, e.g., superconducting material, substrate, and device configuration, make it hard to compare different works \cite{ruf2023effects,ruf2023gate}. As a result, the microscopic origin of the GCS remains controversial. In particular, the role of the leakage current is the key point in the ongoing debate between different studies. Therefore, it will be necessary to understand the microscopic origin behind the GCS effect to assess the reliability of the GCS-based devices to compete with semiconductor technology in various applications.
 
 The low-frequency $1/f^{\alpha}$ noise is widely observed in various nanoelectronic devices \cite{balandin2013low,paladino20141,balogh20211,FALCI20241003}. Investigation of $1/f^{\alpha}$ noise can provide valuable information and a deeper understanding of the transport mechanisms and the dominant sources of fluctuations in different nanodevices \cite{balandin2013low,paladino20141,balogh20211,FALCI20241003}. In addition, noise correlation measurements are very powerful techniques that can be used for probing the cross-correlation between different fluctuating quantities in such systems, which has proven to be successful, for example in the investigation of the bosonic or fermionic, or even the anyonic nature of charge carriers in two-dimensional electron gas \cite{buttiker1992scattering,liu1998quantum,henny1999fermionic,bartolomei2020fractional}. In this work, we address the GCS effect in Al/InAs NWs with $1/f^{\alpha}$ noise measurements for the first time. By carrying out cross-correlation measurements, we show that gate-induced fluctuations in the superconducting NW are directly related to time-dependent leakage current. Furthermore, we performed time-domain measurements to explore the statistics of these fluctuations and the relevant time scales. Finally, based on the detailed analysis of our results, we give a plausible explanation for the microscopic process behind the GCS in the Al/InAs gate-controlled superconducting switches.

\section{Experiments}
\subsection{Device outline and gate influence}
\begin{figure}[tb!]
	\includegraphics[width=\columnwidth]{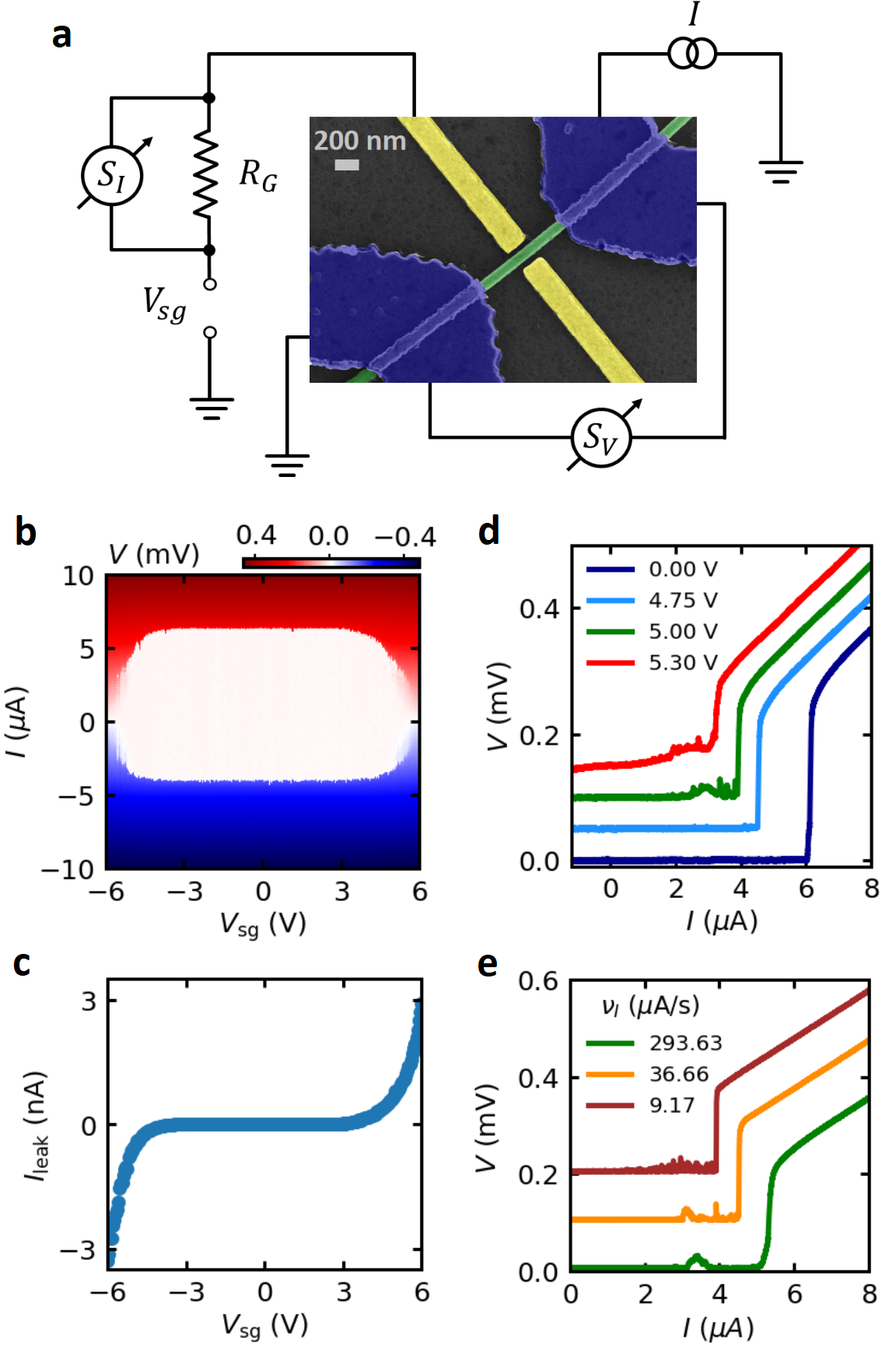}
	\caption{\label{fig:device}\textbf{Device outline and influence of the gate}. (a) A false-colored SEM image and schematic of the external measurement setup. Four Ti/Al contacts (blue) were deposited on top of Al/InAs NW (green) to perform quasi-four probe measurements. Two opposite side gates (yellow) made of Ti/Au were fabricated to investigate the GCS effect in the device. The leakage current is probed by measuring the voltage across the resistor $R_{\mathrm{G}}$=1\,M$\Omega$ connected in series with the gate. (b) $V-I$ characteristics of the NW device as a function of $V_{\mathrm{sg}}$ in which the bias current $I$ is swept from -10 to 10 $\mu A$. The red and blue regions represent the normal state, while the white region corresponds to the superconducting state. (c) The corresponding $I_{\mathrm{leak}}$ as a function of $V_{\mathrm{sg}}$. (d) Selected $V-I$ curves extracted from (a) at different gate voltages in the positive polarity of the bias current and gate voltage. The curves are measured at $\nu_{\mathrm{I}}=293.63\,\mu$A/s (e) $V-I$ curves measured at different current ramp speeds $\nu_{\mathrm{I}}$ at $V_{\mathrm{sg}}=5\,V$. The curves are shifted vertically for clarity. The scattering in the switching current values for the green curves in panels (d) and (e) measured at the same current ramp speed can be attributed to the broad nonthermal switching current distribution induced by the applied gate voltage \cite{puglia2020electrostatic,basset2021gate,ritter2022out,elalaily2023signature}.}
\end{figure}

We have investigated the GCS in Al/InAs NW-based devices with the device configuration shown in the false-colored SEM image in Fig.\,\ref{fig:device}a (see Methods for the fabrication details). Four Ti/Al contacts (blue) were deposited on top of Al/InAs NW (green) for quasi-four probe measurements. Two Ti/Au side gates (yellow) are placed on opposite sides of the NW and used for tuning the supercurrent in the NW.
We first investigated the GCS effect by measuring the $V-I$ characteristics of the device as a function of $V_{\mathrm{sg}}$ in both positive and negative gate polarities, as shown in Fig.\,\ref{fig:device}b, which is similar to earlier observations \cite{elalaily2021gate,elalaily2023signature}. As the bias current is swept from negative to positive polarity, there are two borderlines separating the superconducting state (white region) from the normal state regions (blue and red regions). These lower and upper borderlines represent the gate dependence of the retrapping current $I_{\mathrm{r}}$ and the switching current $I_{\mathrm{SW}}$, respectively. At $V_{\mathrm{sg}}= 0$\,V, the device switches to the normal state at $I_{\mathrm{SW}}\simeq 6\,\mu$A, while it switches from the normal state to the superconducting state at $I_{\mathrm{r}}\simeq 4.3\,\mu$A. As $V_{\mathrm{sg}}$ increases in either positive or negative gate polarity, $I_{\mathrm{SW}}$ is suppressed from a certain threshold gate voltage $V_{\mathrm{sg,onset}}=3.5$\,V until it is fully quenched at the critical gate voltage, $V_{\mathrm{sg,offset}}\approx 6$\,V. At the gate voltages, where $I_{\mathrm{SW}}$ is suppressed, a corresponding enhancement in the leakage current, $I_{\mathrm{leak}}$ is observed (see Fig.\,\ref{fig:device}c), similar to what was reported in Refs.\,\citenum{ritter2021superconducting,elalaily2021gate,ritter2022out,elalaily2023signature,basset2021gate}. Fig.\,\ref{fig:device}d shows selected $V-I$ curves at different gate voltages for the positive polarity of the bias current and gate voltage and separated vertically for clarity. The $V-I$ curve at $V_{\mathrm{sg}}= 0$\,V (blue curve) shows a well-developed zero resistance below $I_{\mathrm{SW}}$, while at higher gate voltages, e.g., at $V_{\mathrm{sg}}= 5$\,V (green curve), finite voltage fluctuations across the NW are observed slightly below $I_{\mathrm{SW}}$ at the bias current values in the interval of [2,4]\,$\mu$A.

Fig.\,\ref{fig:device}e shows the $V-I$ curve measured at $V_{\mathrm{sg}}=5$\,V with different current ramp speeds, $\nu_{\mathrm{I}}$. At $\nu_{\mathrm{I}}=9.17\,\mu$A/s, the $V-I$ curve exhibits noisy region below $I_{\mathrm{SW}}\simeq 6\,\mu$A. As $\nu_{\mathrm{I}}$ is increased up to 293.63\,$\mu$A/s, the NW device becomes less sensitive to these fluctuations, suggesting that the time scale of the fluctuations is the same order of magnitude as the time scale corresponding to the current ramping rate. It also implies that increasing gate voltage does not only suppresses $I_{\mathrm{SW}}$, but also induces a dissipative region in which the voltage across the superconducting NW has dynamical fluctuations.

\subsection{Noise correlation measurements}
Considering the ongoing debate regarding the origin of the GCS effect, it is valuable to investigate the correlation between gate-induced voltage fluctuations in the superconducting NW and the leakage current, both of which are simultaneously monitored in our experimental setup (see schematic of the measurement setup in Fig.\,\ref{fig:device}a and details in Methods). To investigate this, we first studied the low-frequency fluctuations in the leakage current by measuring the power spectral density of the current noise $S_{\mathrm{I}}$ on the resistor $R_{\mathrm{G}}$ as a function of the applied gate voltage $V_{\mathrm{sg}}$ (see the Supporting Information).
\begin{figure}[ht!]
	\includegraphics[width=\columnwidth]{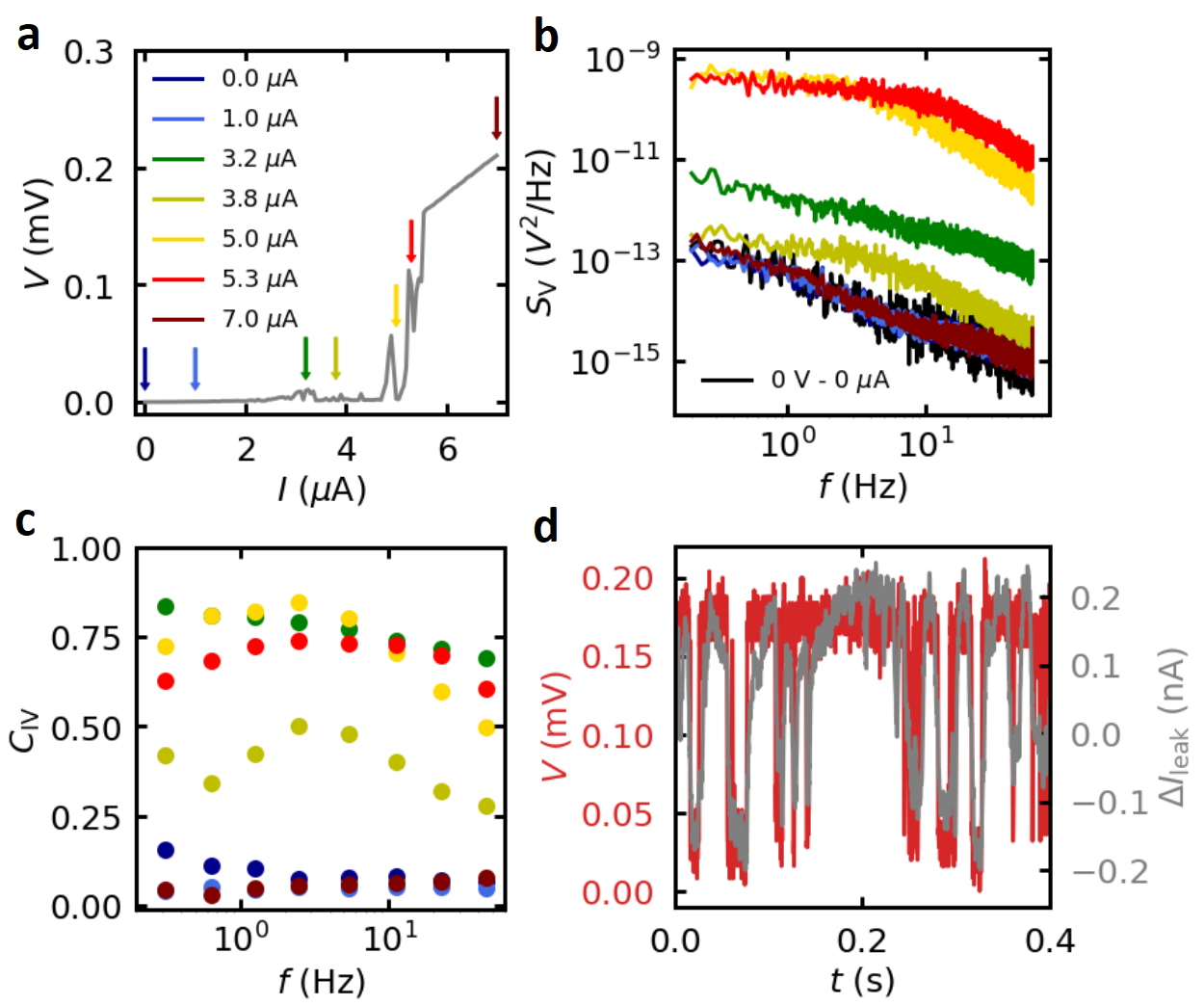}
	\caption{\label{fig:noise_correlation}\textbf{Noise correlation measurements}. (a) The $V-I$ characteristics of the NW device are measured simultaneously with the noise measurements at $V_{\mathrm{sg}}$=5\,V. (b) The noise spectrum of the NW device at different bias current values is pointed out by the colored arrows in panel (a). (c) The Coherence, $C_{{IV}}$ between the leakage current noise and the NW noise spectra presented in panel (b). Every single point represents an average for one octave in $f$. The colors correspond to the current value where they are measured. (d) Time traces for the fluctuations on top of the leakage current (gray time trace) at $V_{\mathrm{sg}}=5$\,V and the voltage fluctuations across the NW device (red time trace) at $I=5.3\,\mu$A.}  
\end{figure}

To investigate the gate-induced voltage fluctuations in the NW device, we fixed $V_{\mathrm{sg}}$ at 5\,V and measured simultaneously the spectral density of the leakage current noise $S_{\mathrm{I}}$ and the induced voltage noise $S_{\mathrm{V}}$ across the NW device as a function of the bias current (see the Supporting Information for complete measurements). Fig.\,\ref{fig:noise_correlation}a shows the $V-I$ curve of the NW device measured in parallel with the noise measurements. The colored arrows point to working points along the $V-I$ curve where the behavior of the measured noise spectra of the NW device will be studied. These points are representative samples from the three characteristic regimes in which the $V-I$ curve exhibits a pure superconducting state (dark and light blue arrows), voltage fluctuations below $I_{\mathrm{SW}}$ (light and dark green, red, and yellow arrows), and a pure normal state (brown arrow) with corresponding noise spectra of $S_{\mathrm{V}}$ shown in Fig.\,\ref{fig:noise_correlation}b. The noise spectra measured in the pure superconducting (dark and light blue curves) have the same amplitude as the background noise originating from the setup (black curve) measured at $I$=\,0\,$\mu$A and $V_{\mathrm{sg}}$=\,0\,V. Similarly, the noise spectrum measured in the normal state (brown curve) also has the same noise level within our experimental resolution. On the other hand, the noise spectra measured in the voltage fluctuation region are more than an order of magnitude larger. Interestingly, the noise spectrum $S_{\mathrm{V}}$ measured at $I=3.2\,\mu$A (dark green curve) has $1/f$ dependence ($\alpha=1$), while the spectra at higher bias current values (light green, red, and yellow curves) have a steeper slope at higher frequencies ($\alpha=2$) resembling a Lorentzian spectrum (see the Supporting Information).

The cross-correlation between the leakage current noise and the induced noise in the NW was investigated in terms of the coherence $C_{{IV}}$ (see Methods). If two signals are fully correlated, coherence has a value of 1, while for uncorrelated signals, it is 0. Fig.\,\ref{fig:noise_correlation}c shows the coherence (averaged per frequency octave) between the two noise signals at different bias current values. The coherence measurements show that it has values of 0.3$-$0.8 for the light and dark green, yellow, and red curves, corresponding to the entire current region where the voltage fluctuation is observed, while it is $C_{{IV}}< 0.2$ for the curves measured at the superconducting and normal states (see the complete measurements in the Supporting Information). This indicates that the voltage fluctuations observed in the superconducting wire are not independent, but they are strongly correlated with the current fluctuation of the leakage between the gate and the superconducting NW. This strong correlation was also confirmed by the time-domain measurements. Fig.\,\ref{fig:noise_correlation}d shows the time-domain measurements for the fluctuations on the top of the average leakage current $\Delta$$I_{\mathrm{leak}}$ (grey curve) and the induced voltage fluctuations in the NW (red curve) at $V_{\mathrm{sg}}=5$\,V and at $I=5.3\,\mu$A. The time traces show that the leakage current fluctuates from a low to a high current state, whereas the voltage across the device changes from a zero-voltage state to a finite-voltage state. This strong correlation is a very robust effect as we have observed a high coherence between the leakage current noise and the induced voltage noise across the superconducting NW in three other devices (see the Supporting Information), which confirms the reproducibility of our results.

\subsection{Time-domain measurements} 
\begin{figure*}[ht!]
	\includegraphics[width=\textwidth]{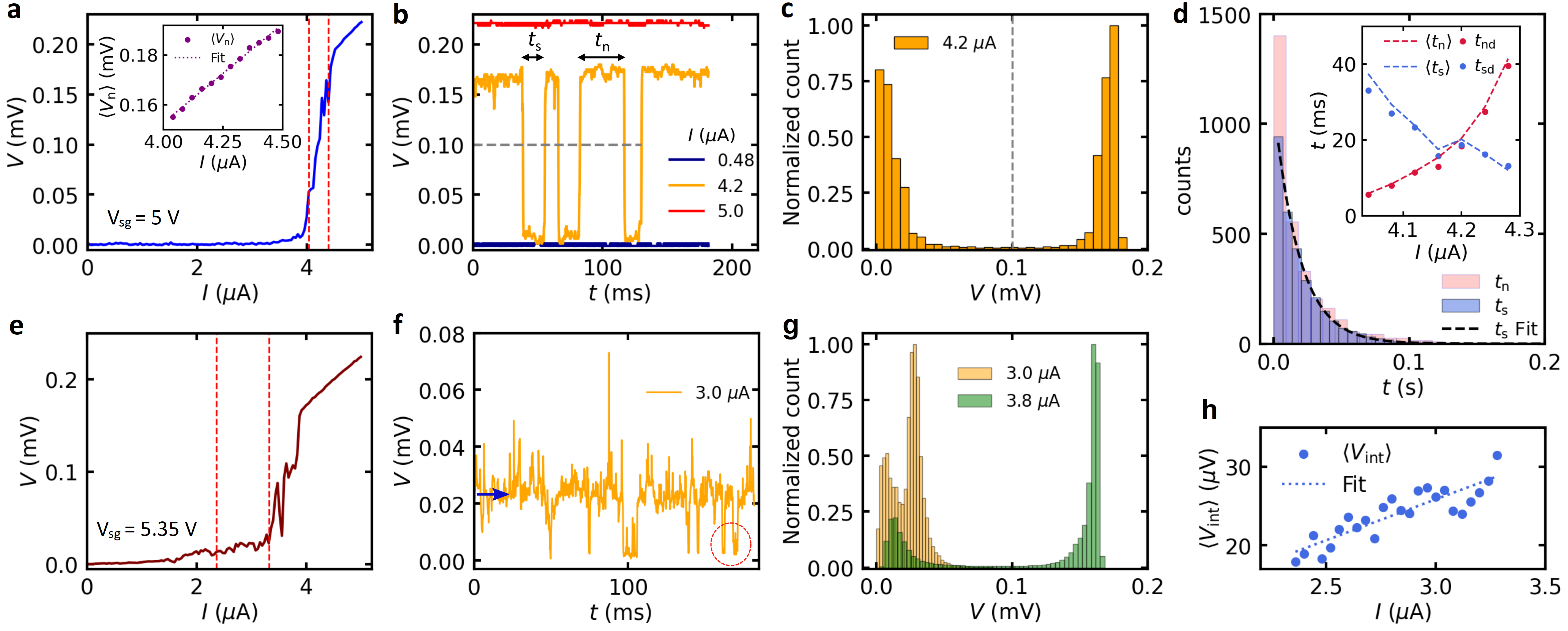}
	\caption{\label{fig:TD_analysis}\textbf{Analysis of the time-domain measurements} (a) The $V-I$ curve of the NW device measured at $V_{\mathrm{sg}}=5$\,V. The inset shows the average high-voltage state $\langle V_{\mathrm{n}} \rangle$ as a function of the bias current in the interval surrounded by the two red dashed lines in panel a. The high-voltage state $\langle V_{\mathrm{n}} \rangle$ is obtained by extracting and averaging the voltage above the threshold at $V=0.1$\,V indicated by the gray dashed lines in panels b and c. The dashed line is the linear fitting of $\langle V_{\mathrm{n}} \rangle$ with a corresponding resistance of $R\approx80\,\Omega$. (b) Representative segments from time traces of the voltage across the NW, $V$, at different bias current values. $t_{\mathrm{s}}$ and $t_{\mathrm{n}}$ represent the lifetimes at which the superconducting NW is in superconducting and normal state, respectively. (c) The probability distribution of the measured voltages along the time traces at $I=4.2\,\mu$A. (d) The distribution of the lifetimes of the high-voltage state $t_{\mathrm{n}}$ (red distribution) and the low-voltage state $t_{\mathrm{s}}$ (blue distribution) extracted from time traces at $V_{\mathrm{sg}}=5\,$V and $I=4.2\,\mu$A. The black dashed curve is a fitting of the probability distribution of the $t_{\mathrm{s}}$ histogram (blue) with an exponential $n=(1/t_{\mathrm{sd}})\exp(-t/t_{\mathrm{sd}})$. The inset shows the decay times $t_{\mathrm{sd}}$ and $t_{\mathrm{nd}}$ (blue and red circles) extracted from the fitting of $t_{\mathrm{s}}$ and $t_{\mathrm{n}}$ distributions and the average lifetimes $\langle t_{\mathrm{s/n}} \rangle$ (blue and red curves) of the two distributions as a function of the bias current. (e) The time average of the $V-I$ curve of the NW device at $V_{\mathrm{sg}}=5.35$\,V. (f) A segment of the time trace of the voltage across the NW, $V$, at $I=3\,\mu$A. The time trace exhibits two different voltage states: a low-voltage state (red circle) and an intermediate-voltage state (blue arrow). (g) The normalized distribution of the measured voltages along the time traces was measured at $I=3\,\mu$A and at $I=3.8\,\mu$A. (h) The average intermediate-voltage state $\langle V_{\mathrm{int}} \rangle$ as a function of the bias current in the interval surrounded by the two red dashed lines in panel (e). The dashed line is the linear fitting of $\langle V_{\mathrm{int}} \rangle$ with a corresponding resistance of $R\approx10\,\Omega$.}
\end{figure*}
To understand the nature of the induced voltage fluctuations in the NW, we measured the voltage across the superconducting NW $V$ as a function of time at a finite gate voltage and different bias current values. Fig.\,\ref{fig:TD_analysis}a shows the time average $V-I$ curve measured at $V_{\mathrm{sg}}=5$\,V. The $V-I$ curve is obtained by averaging a 119\,s-long time trace for each bias current value. The time traces were measured by an ultra-high frequency lock-in amplifier (See Methods) with a sampling rate of 27.5\,kS/s. Fig.\,\ref{fig:TD_analysis}b shows typical time traces of $V$ at different bias current values in which the device exhibits a pure superconducting state (blue trace measured at 0.48\,$\mu$A), a pure normal state (red trace measured at 5\,$\mu$A), and the regime below $I_{\mathrm{SW}}$ (orange trace measured at 4.2\,$\mu$A), where the large voltage fluctuations have been observed. As expected, the time traces measured in the superconducting and normal states show a constant zero-voltage and finite-voltage across the device with time, respectively. On the other hand, the time trace measured in the voltage fluctuation regime shows that the device has voltage fluctuations between two well-defined values, as already seen in Fig.\,\ref{fig:noise_correlation}d. Figure\,\ref{fig:TD_analysis}c presents the distribution of the measured voltages constructed from the time traces and normalized by the maximum bin counts at $I=4.2\,\mu$A. It shows a histogram with two peaks, one with a high probability at 0\,V, which we identify as the superconducting state. To understand the nature of the high voltage state, we set a threshold at $V=0.1$\,mV, as indicated by the gray dashed line between the two voltage states (see Fig.\,\ref{fig:noise_correlation}b,c). The voltage above the threshold was averaged for the whole period, denoted by $\langle V_{\mathrm{n}} \rangle$. This average voltage was extracted for different bias currents in the region marked by the red dashed lines in Fig.\,\ref{fig:TD_analysis}a and plotted as a function of the corresponding bias current (see inset of Fig.\,\ref{fig:TD_analysis}a). The average voltage increases linearly with the bias current with a corresponding resistance of $R\approx80\,\Omega$ which is in the same order of magnitude as the normal state resistance measured at $I>4.5\,\mu$A ($R_{\mathrm{n}}=56\,\Omega$). This indicates that the high voltage state is the dissipative normal state, and the NW in the voltage fluctuation region oscillates between the superconducting and normal states. This is possible since, at $V_{\mathrm{sg}}$=\,5\,V, the voltage fluctuations appear slightly below $I_{\mathrm{r}}$ (see the Supporting Information), therefore, it is expected that if the device switches to the normal state, it will be retrapped to the superconducting state.

 As the next step to analyze the dynamics of the induced fluctuations in the NW, we extracted the lifetimes $t_{\mathrm{s}}$ and $t_{\mathrm{n}}$ in which the NW is in the superconducting and normal state, respectively. This was implemented by extracting the periods at which the voltage across the device is above and below the threshold at 0.1\,V (see Fig.\,\ref{fig:TD_analysis}b). Fig.\,\ref{fig:TD_analysis}d shows the distribution of the current-dependent lifetimes $t_{\mathrm{s}}$ (blue color) and $t_{\mathrm{n}}$ (red color) at $V_{\mathrm{sg}}\,=\,5\,$V and $I=4.2\,\mu$A. The black-dashed curve in Fig.\,\ref{fig:TD_analysis}d is the fitting of the $t_{\mathrm{s}}$ histogram (blue histogram) with an exponential, $n=(1/t_{\mathrm{sd}})\exp(-t/t_{\mathrm{sd}})$ with $t_{\mathrm{sd}}$ as fitting parameter. The fitting shows that the distribution follows an exponential dependence (See also the Supporting Information). The same fitting was implemented for both distributions, for the bias current values in the interval indicated by the red dashed lines in Fig.\,\ref{fig:TD_analysis}a. The inset of Fig.\,\ref{fig:TD_analysis}d presents the extracted decay times $t_{\mathrm{sd}}$ and $t_{\mathrm{nd}}$ (blue and red circles) and the average lifetimes $\langle t_{\mathrm{s/n}} \rangle$ (blue and red curves) for both distributions as a function of the bias current. The nice matching between the two quantities for both lifetimes indicates that the switching of the NW under the influence of the gate in this region resembles the Poisson process \cite{da2011collective}. Furthermore, as the bias current is increased, the lifetime of the NW in the normal state is increased, while it stays for a shorter time in the superconducting state. On the other hand, the distribution of the lifetimes indicates that the gate-induced fluctuations in the NW are slow and can have long time scales.

The analysis of the time-domain measurements for the voltage across the NW at $V_{\mathrm{sg}}$\,=\,5\,V shows well-defined two-level fluctuations between the superconducting and normal states. However, at higher gate voltages, we noticed that the fluctuations of the system were different. Fig.\,\ref{fig:TD_analysis}e presents the time-average $V-I$ curve measured at $V_{\mathrm{sg}}=5.35$\,V. Even though the ramp rate of the bias current is very slow (0.336\,nA/s), we observed that the time-average of the voltage has almost a constant value where the bias current was increased from $2.4\,\mu$A up to $3.28\,\mu$A (the bias current interval indicated by the two red dashed lines in Fig.\,\ref{fig:TD_analysis}e) compared to the same interval in the $V-I$ curve measured at $V_{\mathrm{sg}}=5$\,V shown in Fig.\,\ref{fig:TD_analysis}a. The time trace of the voltage at $I=3\,\mu$A is plotted in Fig.\,\ref{fig:TD_analysis}f, which shows that the system fluctuates between the superconducting state (red circle) and a wide range of different finite voltage values (blue arrow). This can also be observed from the normalized distribution (yellow distribution) of these voltage states shown in Fig.\,\ref{fig:TD_analysis}g. Since the finite-voltage state is not at the same level as the normal-state voltage, we will denote it as the intermediate-voltage state. Fig.\,\ref{fig:TD_analysis}h shows the average voltage of the intermediate state $\langle V_{\mathrm{int}} \rangle$ as a function of the bias current in the interval indicated by the two red dashed lines in Fig.\,\ref{fig:TD_analysis}e. The resistance obtained from fitting the curve with a linear function is $R_{\mathrm{int}}\approx10.4\,\Omega$. which is significantly lower than that of the normal state. As the bias current increases beyond this regime, e.g., at $I=3.8\,\mu$A, the distribution (green distribution in Fig.\,\ref{fig:TD_analysis}g) shows that the system fluctuates between a low-voltage state and the normal state (see also the Supporting Information). These fluctuations resemble those observed at $V_{\mathrm{sg}}=5$\,V, i.e., Fig.\,\ref{fig:TD_analysis}c  (having a resistance of 80\,$\Omega$). 
\section{Discussion}
\begin{figure}[ht!]
	\includegraphics[width=\columnwidth]{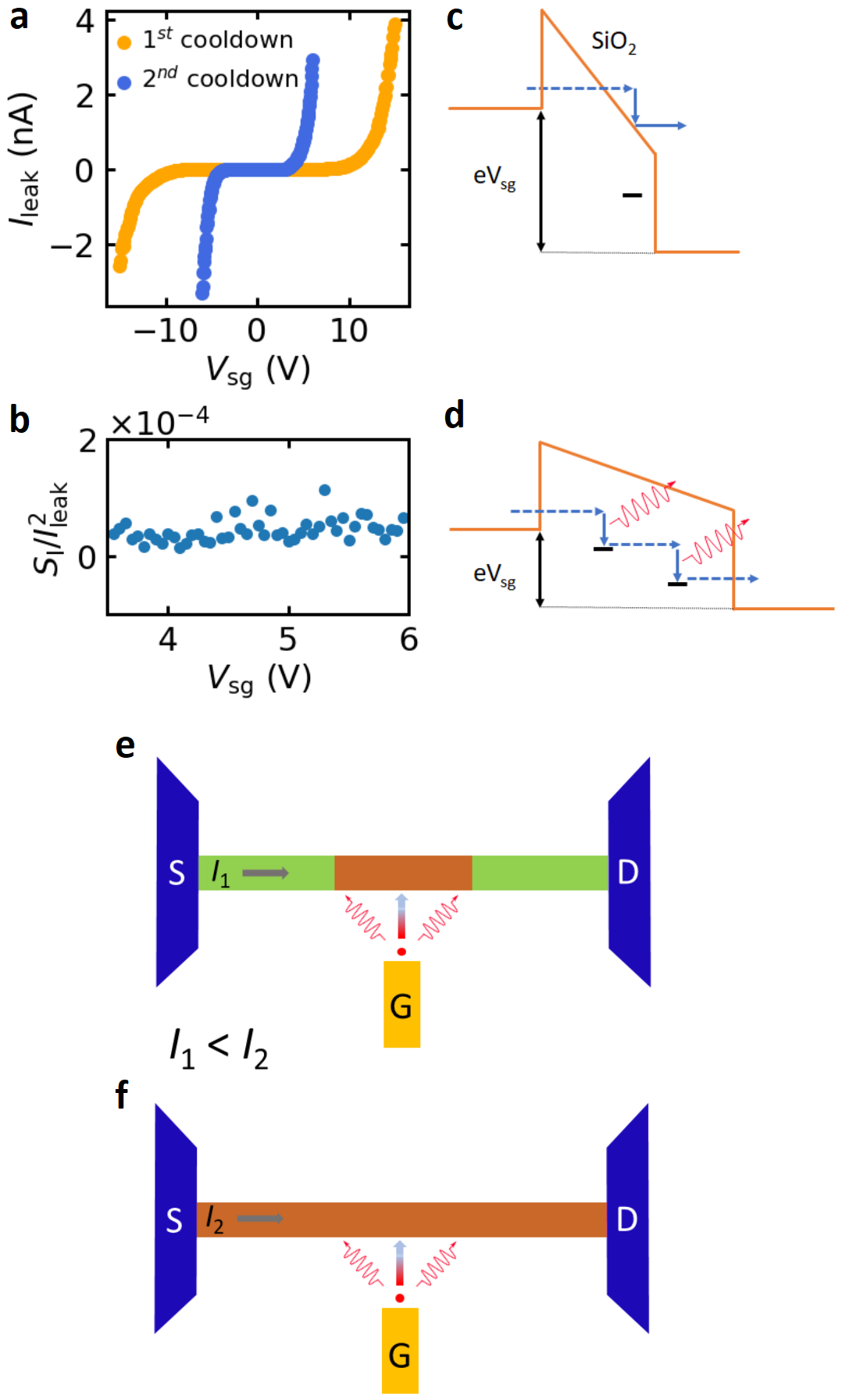}
	\caption{\label{fig:SILC}\textbf{Interpretation of the dynamics of the GCS effect} (a) $I_{\mathrm{leak}}$ as a function of $V_{\mathrm{sg}}$ of the NW device in two different cooldowns. (b) normalized current noise $S_{\mathrm{I}}/I^2_{\mathrm{leak}}$ as a function of $V_{\mathrm{sg}}$ at $f=10\,$Hz, which is consistent with resistance fluctuations. (c) Schematics of generation of the charge traps (black horizontal line) in the oxide layer by the collision of high-energy electrons with the atoms of the oxide \cite{chou1997modeling}. (d) Multiple trap-assisted tunneling. The inelastic tunneling of electrons through the trap states is associated with phonons emission (red arrows). (e) Schematics of the switching mechanism of the superconducting NW (SC) to the normal state (N) by the phonons (red arrows) generated by the inelastic tunneling of electrons (red circle) in the case of low bias current values and (f) high bias current values. For low bias current, only a short segment of the wire switches to normal (the orange part), while for large current, the dissipation in the short segment is enough to bring the entire wire to normal.} 
\end{figure}

In order to understand the origin of the $1/f^{\alpha}$ noise in the leakage current, we should emphasize that the gate dependence of the device has been investigated in two different cooldowns. In the first cooldown, the switching current of the NW device was fully suppressed at $\pm$15\,V applied on the gate (see the Supporting Information) after initial training measurements \cite{elalaily2021gate}. In the second cooldown, this gate voltage window where $I_{\mathrm{SW}}$ is fully suppressed has been reduced to $\pm$6\,V (see Fig.\,\ref{fig:device}b) after initial training. Fig.\,\ref{fig:SILC}a presents the $I_{\mathrm{leak}}$ as a function of $V_{\mathrm{sg}}$ in both cooldowns. The figure shows the leakage current is significantly increased at lower gate voltages in the second cooldown, similar to the trend observed during the initial training as well \cite{elalaily2021gate}. The enhancement of leakage current after applying a higher gate voltage is consistent with the stress-induced leakage current (SILC) that was observed in gate oxides of the MOSFET transistors \cite{maserjian1982behavior,chou1997modeling,alers1998trap}.

Based on Ref.\,\citenum{chou1997modeling}, when a high voltage is applied across the gate oxide in the Fowler-Nordheim regime (see Fig.\,\ref{fig:SILC}c), the high-energy electrons, which tunnel through the oxide potential barrier, suffer collisions with the atoms of the oxide, resulting in damage and hence creating charge traps in the oxide layer. These charge traps serve as intermediate sites, which can increase the leakage current at lower applied voltages via trap-assisted tunneling. In the case of thick oxides, as in our case (the oxide layer between the gate and the NW), multiple traps are assumed to assist in the conduction process \cite{schuler2002physical,ielmini2002modeling}. The generated charge traps can trap electrons and randomly release them with long time constants giving rise to fluctuations in the trap-assisted tunneling and hence the presence of $1/f^{\alpha}$ noise in the $I_{\mathrm{SILC}}$ \cite{ralls1984discrete,alers1998trap}. These time scales are consistent with the time scales of fluctuations observed in our case (see Fig.\,\ref{fig:TD_analysis}d). Furthermore, a characteristic property of SILC is that the normalized current noise $S_{\mathrm{I}}/I^2_{\mathrm{SILC}}$ should be independent of the applied gate voltages \cite{alers1998trap}. Fig.\,\ref{fig:SILC}b shows the $S_{\mathrm{I}}/I^2_{\mathrm{leak}}$ as a function of $V_{\mathrm{sg}}$ at $f=10$\,Hz in the gate voltage window at which $I_{\mathrm{SW}}$ is suppressed. The ratio $S_{\mathrm{I}}/I^2_{\mathrm{leak}}$ is constant with increasing $V_{\mathrm{sg}}$ as expected for SILC process \cite{alers1998trap}. Furthermore, the dominant conduction mechanism in the SILC is inelastic tunneling \cite{takagi1999experimental}, in which electron energy relaxation takes place predominantly with the emission of phonons (see Fig.\,\ref{fig:SILC}d) \cite{sakakibara1997quantitative,jimenez2001physical}. In our case, the emitted phonons can propagate through the substrate, couple with the superconducting NW, and cause the GCS effect \cite{ritter2022out}. In addition, this energy relaxation could also excite surface plasmons, and their decay with photon generation can also contribute to GCS \cite{lambe1976light}. 

Considering the time-domain measurements at 5.35\,V, the appearance of an intermediate-voltage state (see Fig.\,\ref{fig:TD_analysis}f) with a resistance of $R_{\mathrm{int}}=10.4\,\Omega \ll  R_{\mathrm{n}}$ is consistent with a short segment of the NW having been switched to the normal state (see Fig.\,\ref{fig:SILC}e). As $V_{\mathrm{sg}}$ is increased, the leakage current reaches a threshold at which the electrons or the generated phonons can induce phase slips in the superconductor NW at the position where most of the energy is transferred (see orange region) \cite{shah2008inherent,bezryadin2013superconductivity}. Since the bias current is small, the heat dissipated by the phase slip is not enough to switch the whole superconducting NW to its normal state; thus, other sections (green regions) remain superconducting. The ratio $R_{\mathrm{int}}/R_{\mathrm{n}}$ indicates that the NW segment that switched to the normal state has a length of $L_{\mathrm{n}}\simeq160$\,nm, which is consistent with the minimal size of a phase slip center, which has a value of $2\xi\approx156$\,nm (see Methods) \cite{tinkham2004introduction}, where $\xi$ is the superconducting coherence length in the Al shell. However, the leakage current has a strong fluctuation around its main value (see Fig.\,\ref{fig:noise_correlation}d), as a result, the phonon-mediated heat transfer also varies, resulting in a fluctuation in the size of the normal segment and the scattering in the value of $V_{\mathrm{int}}$ (see Fig.\,\ref{fig:TD_analysis}f). We also emphasize that at this gate voltage regime, the switching of the NW to the normal state due to macroscopic quantum tunneling and thermally activated phase slips is negligible compared to the gate-induced phase slips as reported in Refs.\,\citenum{puglia2020electrostatic,elalaily2023signature}.

As the bias current is increased, the heat dissipated by the induced phase slips becomes sufficient to switch the whole NW to the normal state (see Fig.\,\ref{fig:SILC}f). Since $I_{\mathrm{SW}}$ and $I_{\mathrm{r}}$ are almost equal at this gate voltage regime (see the Supporting Information), the NW gets retrapped again to the superconducting state when the fluctuating leakage current gets reduced (see Fig.\,\ref{fig:noise_correlation}d). As a consequence, the voltage across the NW fluctuates between two discrete states (Fig.\,\ref{fig:TD_analysis}c), which induces a Lorenzian spectrum in the power spectral density (see Fig.\,\ref{fig:noise_correlation}b). With further increases in the bias current, the heat dissipated by the phase slips will be enough to switch the entire wire and keep it in its normal state.

The measurements at two different gate voltages are consistent with this picture since at lower gate voltages (e.g., at $V_{\mathrm{sg}}=5\,V$), the induced voltage fluctuations appear at larger bias current values, where the Joule heating in the phase slip center switches the whole NW to the normal state and it retraps back to the superconducting state. At large gate voltages (e.g., at $V_{\mathrm{sg}}=5.35\,V$), the gate-induced fluctuations are increased such that they can induce voltage fluctuations already at low bias current values, but this is only enough to switch a short segment of the NW to the normal state, which results in the appearance of the intermediate voltage state at this elevated gate voltage. At larger bias current values, the two-level fluctuation also shows up, where the NW fluctuates between entire superconducting and normal states.

\section{Conclusions and outlook}
We characterized the GCS effect in Al/InAs NW-based devices by $1/f^{\alpha}$ noise measurements. The noise correlation measurements show a large correlation between the leakage current noise and the induced voltage noise in the superconducting NW. Our findings suggest that the GCS effect originated from phonons generated by inelastic tunneling of the electrons through the trap states created by stressing the oxide layer between the gate and the NW under a high electric field, which is qualitatively consistent with the microscopic picture proposed by Ref.\,\cite{ritter2022out}.

Moreover, our results show that even an individual resistance fluctuator can control the switching of the supercurrent in the nanowire. This is an important finding, which means that even a very small change in the inelastic tunneling current at a critical spot can lead to abrupt switching of the supercurrent. Identification of such a critical spot would make these systems very promising elements for fast superconducting electronics. These spots are likely in close contact with the nanowire, where the influence of the generated phonons is maximized. Furthermore, besides phonon emission in inelastic tunneling, there should also be photon emission, which could also contribute to the switching of the superconductor. Such photons could be directly verified using low-temperature optical emission experiments.

Considering superconducting gating for fast logic circuits, the speed of switching back to the superconducting state is expected to be the limiting factor since it is governed by thermal processes and their intrinsic speed. With Al superconductor below 1\,K the thermal response speeds are limited to frequencies $f \leq 1$\,GHz owing to low superconducting transition temperature. However, by using materials with large critical temperatures such as NbTiN \cite{machhadani2019improvement}, much larger thermal speeds could be achieved. Presuming that the operation temperature can be increased from $\sim 1$ to 5\,K, electronic thermal cooling will be replaced by electron-phonon heat transfer \cite{gershenzon1990electron}, and the operation speed can be increased by two orders of magnitude, which is suitable for the implementation of fast logic gates. Furthermore, the GCS effect can be useful in various quantum circuits as a fast gate tunable superconducting-normal switch. This could find application e.g., in SQUID-based CPR measurements \cite{della2007measurement} to switch off the reference junction, or such a unit can be used as a heat switch as well.

\section*{Methods}
The devices were fabricated by depositing InAs NWs with a 20\,nm-thick full Al shell layer (green) \cite{krogstrup2015epitaxy} on an undoped \text{Si} wafer with a 290\,nm-thick oxide layer. After NWs deposition, four Ti/Al contacts (blue) with a thickness of 10/80\,nm and two opposite side gates (yellow) made of Ti/Au with a thickness of 7/33\,nm were fabricated in two separate electron beam lithography (EBL) steps. For the device shown in Fig.\,\ref{fig:device}a, the separation between the Al contacts is $\simeq 870$\,nm. The distance between the upper gate electrode and the NW is $\simeq 50$\,nm, while for the lower gate electrode, it is $\simeq70$\,nm. All the GCS-based measurements on this particular device were performed by applying a voltage $V_{\mathrm{sg}}$ to the upper gate electrode. 

The $V-I$ characteristics of the device were measured using a quasi-4-probe method in which the current is ramped linearly from negative to positive values by an engineered voltage signal generated by an arbitrary wave function generator (KEYSIGHT 33600A) with a series resistor of 50\,k$\Omega$, while the voltage across the device is measured with a differential low-noise voltage amplifier and a Z\"urich instruments (UHF-LI 600\,MHz) used as an oscilloscope. The ramping rate of the bias current can be controlled by adjusting the sampling frequency of the waveform generator.

The leakage current noise and the voltage noise in the NW device were probed by measuring the noise signals at the gate resistor $R_{\mathrm{G}}$ and the NW device, respectively. The noise signals were measured by a 2-channel spectrum analyzer (HP89410A). The obtained noise spectra were averaged 20 times. The cross-spectrum and the coherence between the two signals were calculated internally by the spectrum analyzer.

The high-resolution time traces for the voltage fluctuations in the NW device were measured by the quasi-4-probe measurements, whereas the bias current is kept fixed and the voltage across the device is measured as a function of time by the Z\"urich instruments. On the other hand, the time traces in which both the two signals (leakage current and NW fluctuations) were measured simultaneously were probed by a multi-channel oscilloscope (Tektronix TDS 2014B). 

The coherence $C_{{IV}}$ between two signals $X_{{I}}(t)$ and $Y_{{V}}(t)$ is defined at a given frequency $f$ as follows: \cite{carter1987coherence}
\begin{equation}
    C_{{IV}}(f)={\frac{|S_{{IV}}(f)|^{2}}{S_{{I}}(f)S_{{V}}(f)}},
\end{equation}
where, 
\begin{equation}
   {\displaystyle {\begin{aligned}S_{IV}(f)&=\int _{-\infty }^{\infty }E\left[X_{{I}}(t) Y_{{V}}(t+\tau_{l})\right] e^{-i2\pi f\tau_{l} }d\tau_{l}. \end{aligned}}}
\end{equation}
is the cross-spectrum between the two signals, $\tau_{l}$ is the time delay between the two signals, $E$ denotes the the expectation value. $S_{{I}}$ and $S_{{V}}$ are the corresponding power spectral densities of the two signals. $C_{{IV}}$ takes values between 0 and 1, and if the value of the coherence is close to 1, the two signals are strongly correlated. Conversely, the two signals are uncorrelated if the value of the coherence is close to 0 \cite{carter1987coherence}.

The coherence length $\xi$ was calculated by using $\xi=\sqrt{\hbar l/R_{\mathrm{n}}wtN_{\mathrm{f}}e^2\Delta_{\mathrm{0}}}$ \cite{de2018metallic}, where $l=870$\,nm, $w=100$\,nm and $t=20$\,nm are the length, width, and thickness of the NW device, respectively, $N_{\mathrm{f}}=2.15\cdot10^{47}\,\text{J}^{-1}\text{m}^{-3}$ is the density of states at Fermi level, and $\Delta_{\mathrm{0}}=1.76 k_{\mathrm{B}}T_{\mathrm{c}}$, $T_{\mathrm{c}}$= 1\,K.

\section*{Author contributions}
 T.E. and M.B. fabricated the devices, T.E., I.L., and A.S. performed the measurements T.E., M.B., G.F., and L. K. did the data analysis. T.K. and J.N. developed the nanowires. All authors discussed the results and worked on the manuscript. P.M., P.H., and S.C. guided the project.

\section*{Data availability}
The data in this publication are available in numerical form at: \url{https://zenodo.org/records/11004848}. Time domain data is available upon request. 
 
\section*{Acknowledgments}
This work was funded by the EU’s Horizon 2020 research and innovation program under grant Supergate (964398), COST Action CA 21144 superqumap, ERC Twistrain, the FET Open AndQC grant (828948), OTKA K138433 and EIC Pathfinder Challenge QuKiT (101115315), BME and NKFIH, the grant TKP2021-NVA-02 of NKFIH and VEKOP 2.3.3-15-2017-00015. The work was also supported by Quantum Information National Laboratory of Hungary (Grant No. 2022-2.1.1-NL-2022-00004), the ÚNKP-23-3-I-BME-238 New National Excellence Program of the Ministry for Culture and Innovation from the source of the National Research, Development and Innovation Fund, the Academy of Finland projects 341913 (EFT), and 312295 \&  352926 (CoE, Quantum Technology Finland). The research leading to these results has received funding from the European Union’s Horizon 2020 Research and Innovation Programme, under Grant Agreement no 824109 (EMP). This research made use of the OtaNano – Low-Temperature Laboratory infrastructure of Aalto University. This research was supported by the Carlsberg Foundation, and the Danish National Research Foundation (DNRF 101).

\section*{Competing interests}
The authors declare no competing interests.

\bibliography{references}

\end{document}


\title{Switching dynamics in Al/InAs nanowire-based gate-controlled superconducting switch}

\author{Tosson Elalaily}
\affiliation{Department of Physics, Institute of Physics, Budapest University of Technology and Economics, M\"uegyetem rkp. 3., H-1111 Budapest, Hungary}
\affiliation{MTA-BME Superconducting Nanoelectronics Momentum Research Group, M\"uegyetem rkp. 3., H-1111 Budapest, Hungary}
\affiliation{Department of Physics, Faculty of Science, Tanta University, Al-Geish St., 31527 Tanta, Gharbia, Egypt}

\author{Martin Berke}
\affiliation{Department of Physics, Institute of Physics, Budapest University of Technology and Economics, M\"uegyetem rkp. 3., H-1111 Budapest, Hungary}
\affiliation{MTA-BME Superconducting Nanoelectronics Momentum Research Group, M\"uegyetem rkp. 3., H-1111 Budapest, Hungary}

\author{Ilari Lilja}
\affiliation{Low-Temperature Laboratory, Department of Applied Physics, Aalto University School of Science, P.O. Box 15100, Fi-00076 Aalto, Finland}
\affiliation{QTF Centre of Excellence, Department of Applied Physics, Aalto University, Fi-00076 Aalto, Finland}

\author{Alexander Savin}
\affiliation{QTF Centre of Excellence, Department of Applied Physics, Aalto University, Fi-00076 Aalto, Finland}

\author{Gerg\H{o} F\"ul\"op}
\affiliation{Department of Physics, Institute of Physics, Budapest University of Technology and Economics, M\"uegyetem rkp. 3., H-1111 Budapest, Hungary}
\affiliation{MTA-BME Superconducting Nanoelectronics Momentum Research Group, M\"uegyetem rkp. 3., H-1111 Budapest, Hungary}

\author{L\H{o}rinc Kup\'as}
\affiliation{Department of Physics, Institute of Physics, Budapest University of Technology and Economics, M\"uegyetem rkp. 3., H-1111 Budapest, Hungary}
\affiliation{MTA-BME Superconducting Nanoelectronics Momentum Research Group, M\"uegyetem rkp. 3., H-1111 Budapest, Hungary}

\author{Thomas Kanne}
 \affiliation{Center for Quantum Devices and Nano-Science Center, Niels Bohr Institute, University of Copenhagen, Universitetsparken 5, DK-2100, Copenhagen, Denmark}
 
\author{Jesper Nygård}
 \affiliation{Center for Quantum Devices and Nano-Science Center, Niels Bohr Institute, University of Copenhagen, Universitetsparken 5, DK-2100, Copenhagen, Denmark}

\author{P\'eter Makk}
\affiliation{Department of Physics, Institute of Physics, Budapest University of Technology and Economics, M\"uegyetem rkp. 3., H-1111 Budapest, Hungary}
\affiliation{MTA-BME Correlated van der Waals Structures Momentum Research Group, M\"uegyetem rkp. 3., H-1111 Budapest, Hungary}

\author{Pertti Hakonen}
\affiliation{Low-Temperature Laboratory, Department of Applied Physics, Aalto University School of Science, P.O. Box 15100, Fi-00076 Aalto, Finland}
\affiliation{QTF Centre of Excellence, Department of Applied Physics, Aalto University, Fi-00076 Aalto, Finland}

\author{Szabolcs Csonka}
\affiliation{Department of Physics, Institute of Physics, Budapest University of Technology and Economics, M\"uegyetem rkp. 3., H-1111 Budapest, Hungary}
\affiliation{MTA-BME Superconducting Nanoelectronics Momentum Research Group, M\"uegyetem rkp. 3., H-1111 Budapest, Hungary}
\affiliation{Institute of Technical Physics and Materials Science, Center for Energy Research, Konkoly-Thege Mikl\'os \'ut 29-33., 1121, Budapest, Hungary}
\maketitle

\subsection{Investigation of the leakage current noise}
\label{sssec:leakage current exponent} 
\begin{figure}[H]
\begin{center}
        \includegraphics[width=0.7\textwidth,width=\columnwidth]{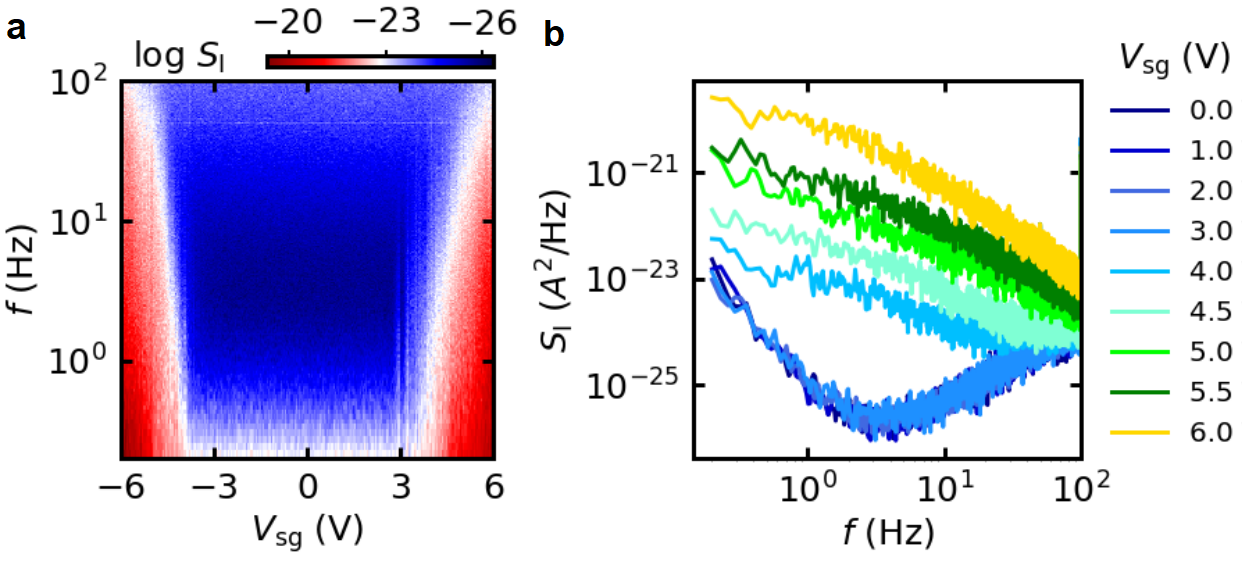}
	\caption{\label{fig:gating_3_leakage noise} Investigation of the leakage current noise. (a) The logarithm of power spectral density of leakage current noise on the plane spanned by $V_{\mathrm{sg}}$ and log $f$. (b) Noise spectra for different gate voltage values extracted from the measurement plotted in panel (a).}
\end{center}
\end{figure}

To investigate the fluctuations in the leakage current, we measured the power spectral density of the current noise $S_{\mathrm{I}}$ on the preresistor $R_{\mathrm{G}}$ as a function of the applied gate voltage $V_{\mathrm{sg}}$ in both negative and positive polarities, as shown in SFig.\,\ref{fig:gating_3_leakage noise}a. The spectral density of the noise was measured only up to 100\,Hz due to the limited bandwidth of the measurement setup defined by the preresistor $R_{\mathrm{G}}=1$\,M$\Omega$ and the capacitance between the wiring of the setup $C_{\mathrm{W}}\approx2$\,nF. As $V_{\mathrm{sg}}<V_{\mathrm{sg,onset}}$ ($I_{\mathrm{leak}}=0$\,A), no change in $S_{\mathrm{I}}$ with $V_{\mathrm{sg}}$ is observed (blue region). Beyond $V_{\mathrm{sg,onset}}$ ($V_{\mathrm{sg,onset}}=3.5$\,V), the leakage current increases exponentially with $V_{\mathrm{sg}}$ (see Fig.\,1c in the main text), and a corresponding enhancement in $S_{\mathrm{I}}$ level is observed (red regions). SFig.\,\ref{fig:gating_3_leakage noise}b shows the noise frequency spectrum for different gate voltage values plotted in the log-log scale. The frequency spectra of the noise at low gate voltages ($V_{\mathrm{sg}}<V_{\mathrm{sg,onset}}$) have the same level and shape as the background spectrum at $V_{\mathrm{sg}}=0$\,V, while at higher gate voltages ($V_{\mathrm{sg}}>V_{\mathrm{sg,onset}}$) the noise level increased and the spectra have $1/f^{\alpha}$ dependence. 

\begin{figure}[H]
\begin{center}
	\includegraphics[width=0.9\columnwidth]{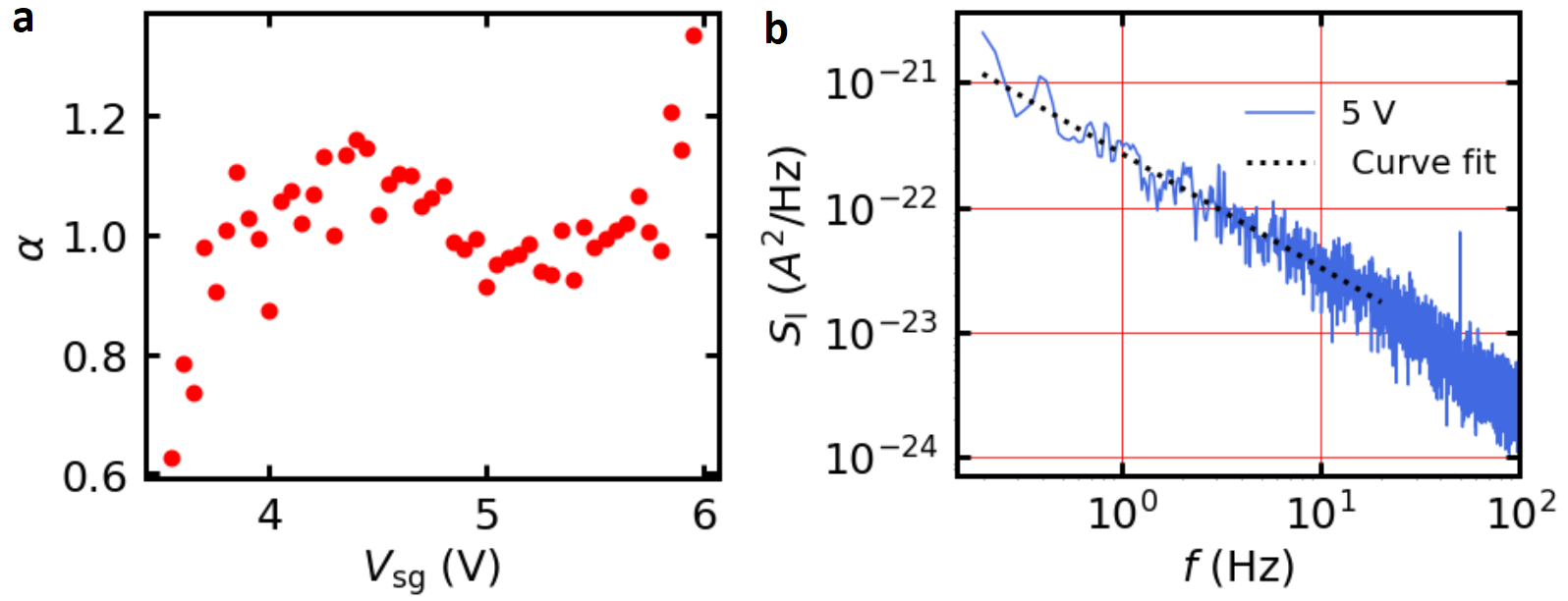}
	\caption{\label{fig:exponent at different gate voltages} Fitting of the leakage current noise. (a) The frequency scaling exponent $\alpha$ of the leakage current noise as a function of the gate voltage in the positive gate polarity. (b) Fitting (black curve) of the frequency spectrum at $V_{\mathrm{sg}}$=5\,V (blue curve).} 
\end{center} 
\end{figure}

Fig.\,\ref{fig:exponent at different gate voltages}a shows the scaling exponent $\alpha$ of $S_{\mathrm{I}}$ as a function of the gate voltage in the positive gate polarity. The exponent is extracted by fitting the spectra at different gate voltages in the frequency range of 0.2-20\,Hz. The gate dependence of $\alpha$ shows that it has a value of $\simeq\,1$ at the gate voltages, where $I_{\mathrm{SW}}$ gets suppressed ($V_{\mathrm{sg,onset}}=3.5-6$\,V). An example of fitting the frequency spectrum at $V_{\mathrm{sg}}$=5\,V is shown in Fig.\,\ref{fig:exponent at different gate voltages}b.
 
\subsection{Investigation of the NW noise} 
\label{sssec:Investigation of the leakage current noise and the NW noise}
\begin{figure}[ht!]
\begin{center}
        \includegraphics[width=0.7\textwidth,width=0.75\columnwidth]{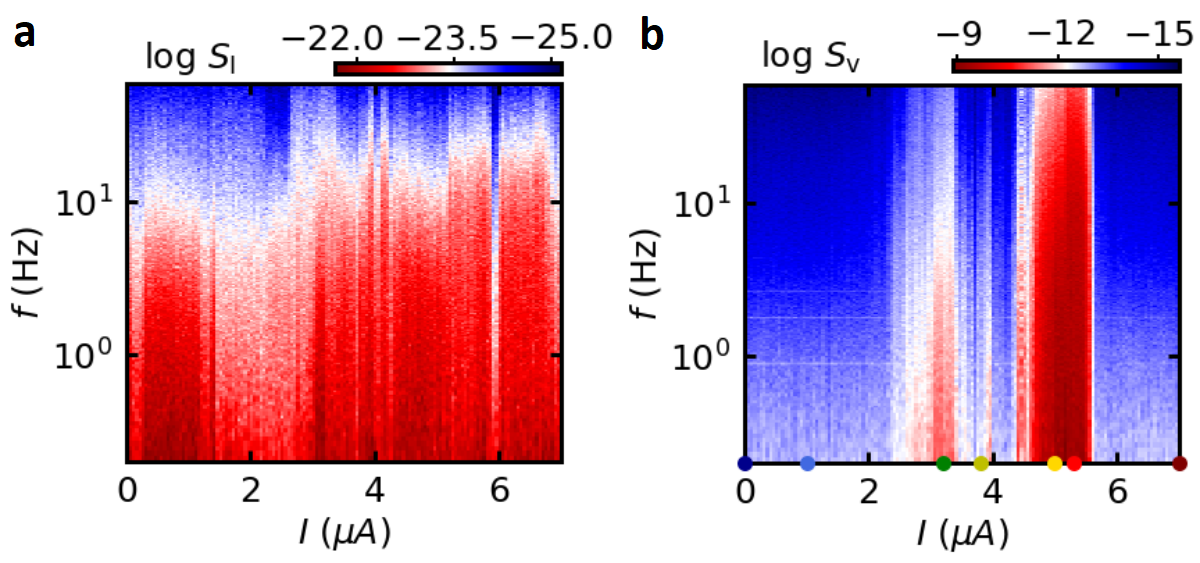}
	\caption{\label{fig:gating_3_wire noise} Investigation of the induced fluctuations in the NW device. (a) The logarithm of power spectral density of leakage current noise $S_{\mathrm{I}}$ and (b) of the NW noise $S_{\mathrm{V}}$ on the plane spanned by $I$ and log $f$. The two measurements were performed simultaneously at a fixed gate voltage, $V_{\mathrm{sg}}$= 5\,V.}
\end{center}
\end{figure}

SFigs.\,\ref{fig:gating_3_wire noise}a and b show the power spectral density of the leakage current noise $S_{\mathrm{I}}$ and the induced voltage noise $S_{\mathrm{V}}$ across the NW device as a function of bias current that were measured simultaneously at $V_{\mathrm{sg}}$ = 5\,V, respectively. As the measurement is performed at constant gate voltage, no significant change in the $S_{\mathrm{I}}$ level is observed (see SFig.\,\ref{fig:gating_3_wire noise}a). However, a large enhancement in the magnitude of $S_{\mathrm{V}}$ at some regions in the bias current below $I_{\mathrm{SW}}$ is observed (see SFig.\,\ref{fig:gating_3_wire noise}b). The colored circles are positioned at the bias currents corresponding to the noise spectra presented in Fig.\,2b in the main text.

\subsection{Fitting of the NW noise spectra} 
\label{sssec:Fitting of the NW noise spectra}
\begin{figure}[H]
\begin{center}
        \includegraphics[width=0.6\textwidth,width=0.6\columnwidth]{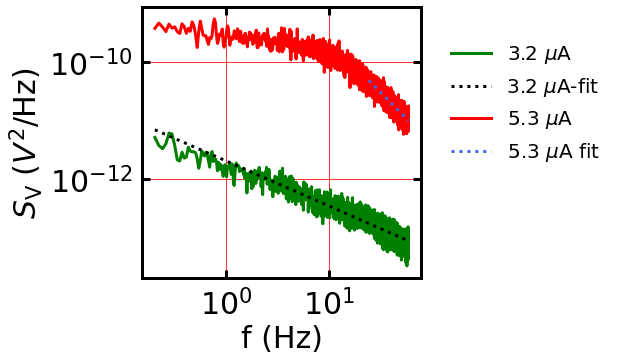}
	\caption{\label{fig:wire noise} Fitting the noise spectra of the NW device $S_{\mathrm{V}}$ measured at $V_{\mathrm{sg}}$= 5\,V at $I=3.2\,\mu$A and $I=5.3\,\mu$A.}
\end{center}
\end{figure}

As we discussed in the main text, the spectra of the voltage noise $S_{\mathrm{V}}$ at $V_{\mathrm{sg}}$= 5\,V (see Fig.\,2b) have almost $1/f$ dependence at low bias current values, while they are Lorentzian at higher bias current values. SFig.\,\ref{fig:wire noise} shows the fitting of the noise spectra at $I=3.2\,\mu$A (green curve) and at $I=5.2\,\mu$A (red curve). The fitting shows that the noise spectrum measured at $I=3.2\,\mu$A has a scaling exponent of $\alpha \simeq 0.8$, while the fitting of the spectrum measured at $I=5.2\,\mu$A at higher frequencies shows a steeper slope with $\alpha \simeq 1.78\pm0.02$. 

\subsection{Cross-spectrum and coherence as a function of the bias current} 
\label{sssec:Cross-spectrum and coherence as a function of the bias current}
\begin{figure}[ht!]
\begin{center}
	\includegraphics[width=0.85\columnwidth]{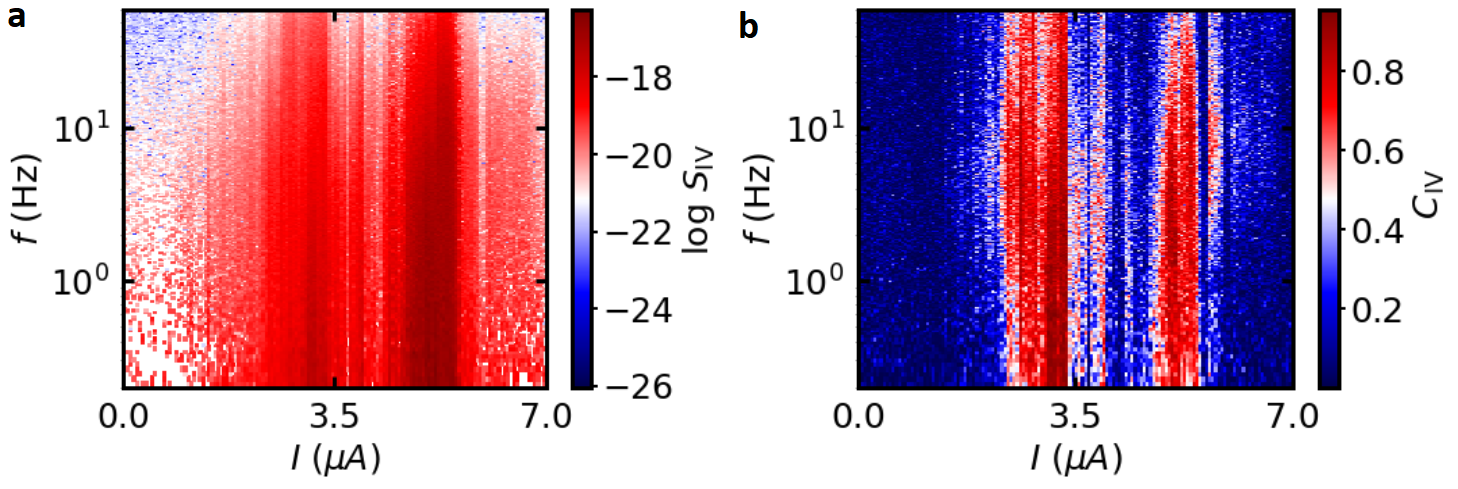}
	\caption{\label{fig:cross-spectrum and coherence} (a) logarithm of the cross-spectrum as a function of the frequency and the bias current at $V_{\mathrm{sg}}=5$\,V (b) The corresponding coherence as a function of the frequency and the bias current.}
\end{center} 
\end{figure}

We showed in Fig.\,2c in the main text the Coherence $C_{\mathrm{IV}}$ between the leakage current noise and the voltage noise across the NW device only at some selected current bias values. Fig.\,\ref{fig:cross-spectrum and coherence}a and b show both the cross-spectrum $S_{\mathrm{IV}}$ and $C_{\mathrm{IV}}$ as a function of frequency for all measured values, respectively.

\subsection{Retrapping current at $V_{\mathrm{sg}}$= 5\,V} 
\label{sssec:Retrapping current}
\begin{figure}[ht!]
\begin{center}
	\includegraphics[width=0.55\columnwidth]{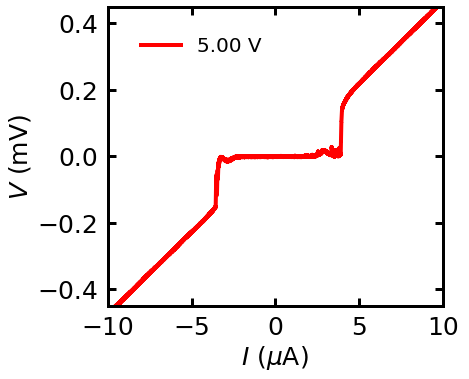}
	\caption{\label{fig:retrapping current} Full $I-V$ curve at $V_{\mathrm{sg}}=5$\,V. The curve shows that the voltage fluctuation appears below both $I_{\mathrm{SW}}$ and $I_{\mathrm{r}}$ in the positive and negative polarity of the bias current, respectively. }
\end{center} 
\end{figure}

In the main text, we presented the $I-V$ characteristics of the NW device at $V_{\mathrm{sg}}$= 5\,V (Fig. 1d) only with the positive polarity of the bias current, in which the voltage fluctuations appeared slightly below $I_{\mathrm{SW}}$. SFig.\,\ref{fig:retrapping current} shows the full $I-V$ curve with positive and negative bias current polarities. Interestingly, at this gate voltage regime, the values of $I_{\mathrm{SW}}$ and $I_{\mathrm{r}}$ are very close to each other, thus the voltage fluctuation also appears below $I_{\mathrm{r}}$ as the device is switched from the normal to the superconducting state. Therefore, it is expected that if the NW device switches to normal under the influence of gate fluctuations, it can be returned to the superconducting state. 

\subsection{Investigation of exponential behavior of the lifetime distributions}
\label{sssec:Investigation of exponential behavior of the lifetime distributions} 

\begin{figure}[ht!]
\begin{center}
	\includegraphics[width=0.5\columnwidth]{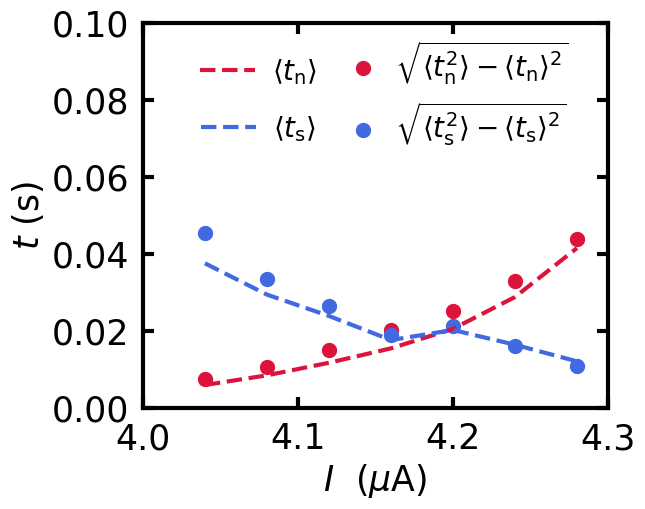}
	\caption{\label{fig:exponential} The mean value $\langle t_{\mathrm{s/n}} \rangle$ (blue/red curves) and the standard deviation $\sqrt{\langle t_{\mathrm{s/n}}^2 \rangle- {\langle t_{\mathrm{s/n}} \rangle^2}}$ (blue/red circles) of the lifetime distributions $t_{\mathrm{s}}$ and $t_{\mathrm{n}}$ as a function of bias current.}
\end{center} 
\end{figure}

As an initial step to investigate the exponential behavior of the lifetime distributions $t_{\mathrm{s}}$ and $t_{\mathrm{n}}$, we extracted the mean value $\langle t_{\mathrm{s/n}} \rangle$ and the standard deviation $\sqrt{\langle t_{\mathrm{s/n}}^2 \rangle- {\langle t_{\mathrm{s/n}} \rangle^2}}$ of both distributions for different values of the bias current as shown in SFig.\,\ref{fig:exponential}. The strong matching between the two quantities, along with the well-fitting of the distributions with an exponential as discussed in the main text, confirms that the lifetime distributions $t_{\mathrm{s}}$ and $t_{\mathrm{n}}$ are exponential distributions and satisfy one of the conditions for a Poisson process \cite{da2011collective}.

\subsection{Noise correlation measurements of other devices} 
\label{sssec:confirmation of the noise correlation}
\begin{figure}[ht!]
\begin{center}
        \includegraphics[width=0.75\textwidth,width=0.75\columnwidth]{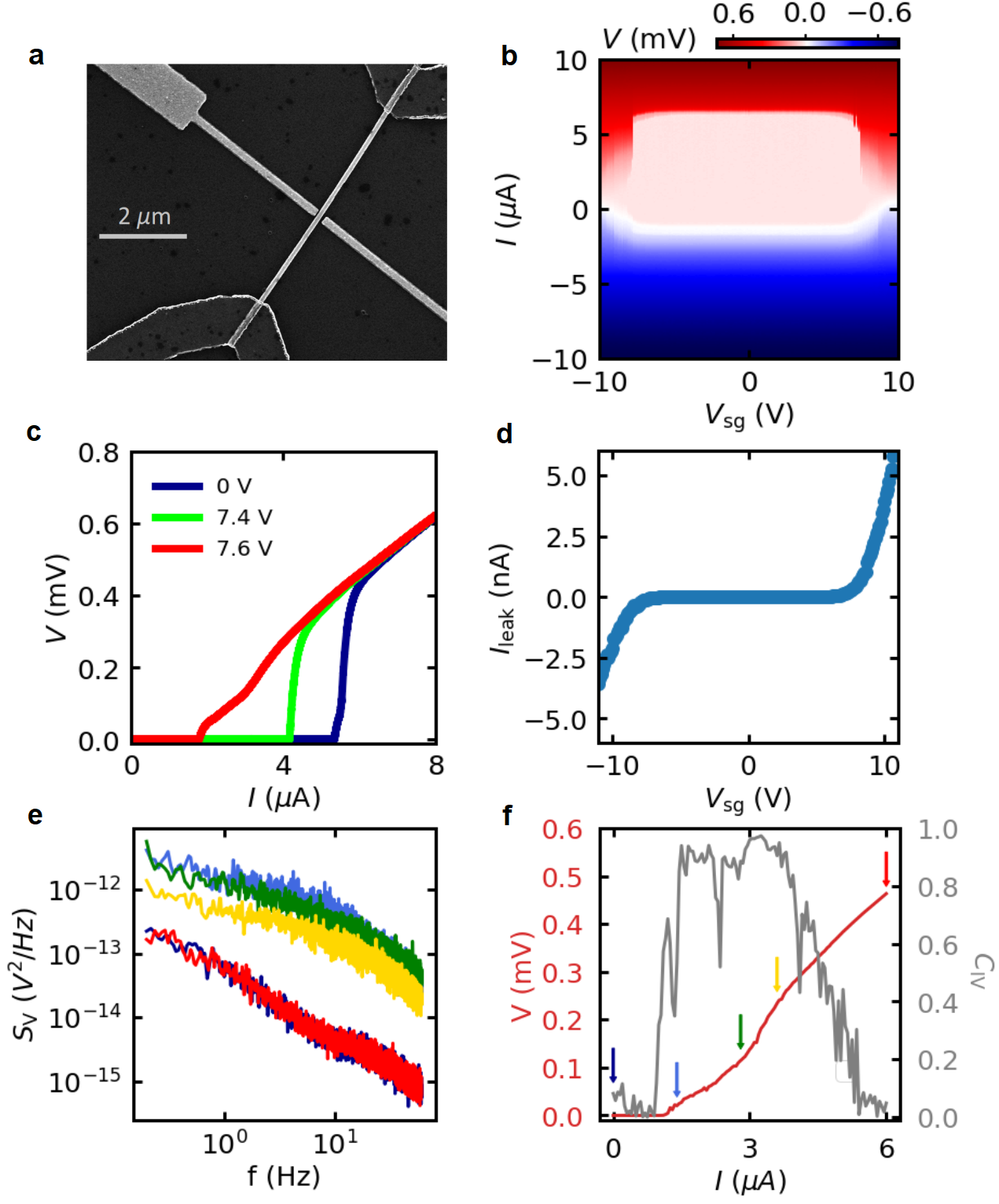}
	\caption{\label{fig:device_2} Investigation of the GCS and the noise correlations in device B ($L_{\mathrm{NW}}$=5.16\,$\mu$m). (a) SEM image of the investigated device. (b) $I-V$ characteristics of the NW device as a function of $V_{\mathrm{sg}}$. (c) Some selected $I-V$ curves are extracted from (b) at different gate voltages in the positive gate polarity. (d) The corresponding $I_{\mathrm{leak}}$ as a function of $V_{\mathrm{sg}}$. The $I-V$ curve measured at $V_{\mathrm{sg}}=7.6$\,V (red curve) shows the presence of a finite voltage state. (e) The NW noise spectra measured at $V_{\mathrm{sg}}=7.6$\,V at different bias current values indicated by the arrows in panel (f) that match the same color. (f) The $I-V$ curve that measured simultaneously with noise measurements and the corresponding coherence (at 3.35\,Hz) between the leakage current noise and the NW noise.}
\end{center}
\end{figure}

\begin{figure}[ht!]
\begin{center}
        \includegraphics[width=0.75\textwidth,width=0.75\columnwidth]{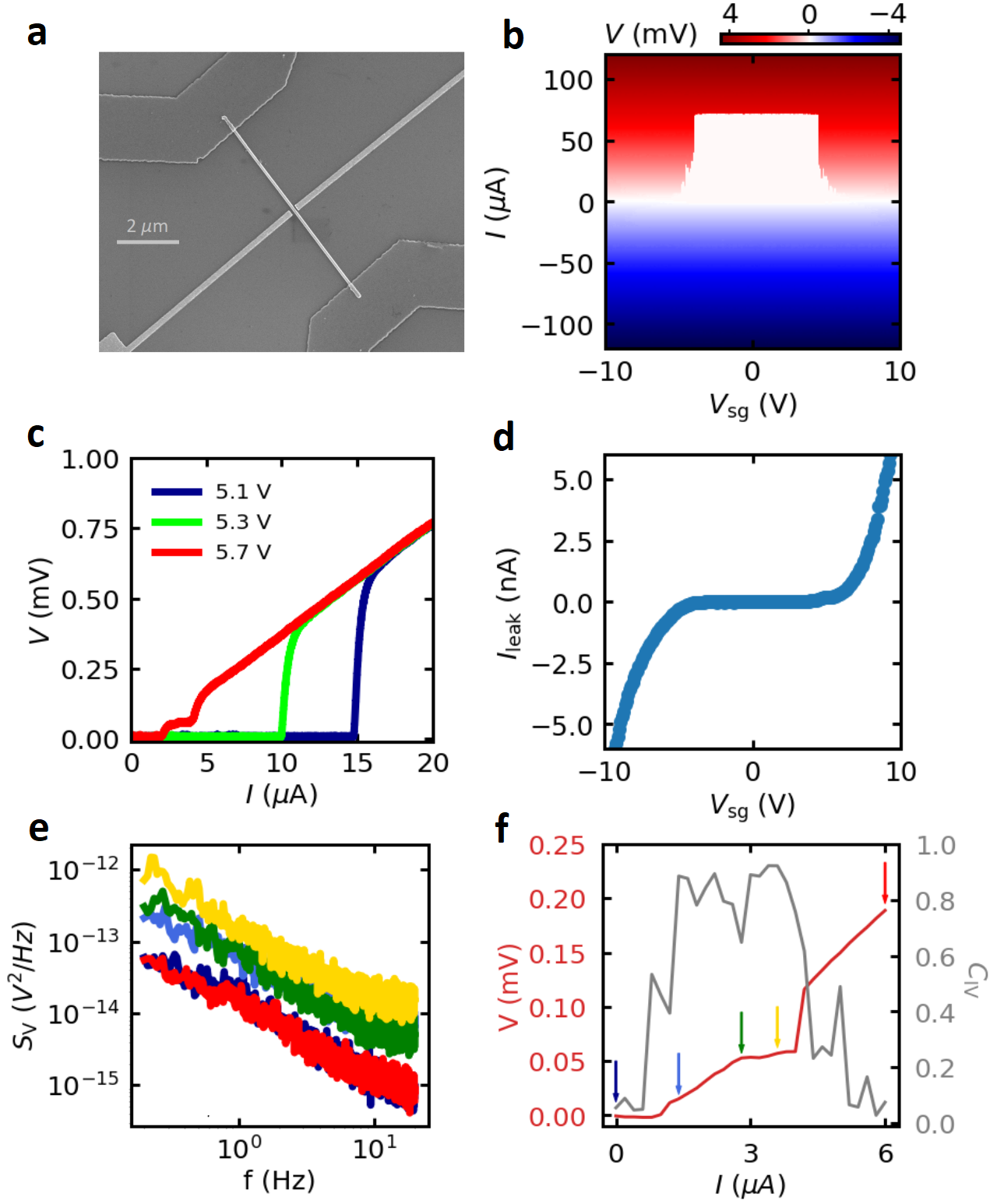}
	\caption{\label{fig:device_3} Investigation of the GCS and the noise correlations in device C ($L_{\mathrm{NW}}$=6.2\,$\mu$m). (a) SEM image of the investigated device. (b) $I-V$ characteristics of the NW device as a function of $V_{\mathrm{sg}}$. (c) Some selected $I-V$ curves are extracted from (b) at different gate voltages in the positive gate polarity. The $I-V$ curve measured at $V_{\mathrm{sg}}=5.7$\,V (red curve) shows the presence of the voltage state. (d) The corresponding $I_{\mathrm{leak}}$ as a function of $V_{\mathrm{sg}}$. (e) The NW noise spectra measured at $V_{\mathrm{sg}}=5.7$\,V at different bias current values indicated by the arrows in panel (f) that match the same color. (f) The $I-V$ curve that measured simultaneously with noise measurements and the corresponding coherence (at 3.35\,Hz) between the leakage current noise and the NW noise.}
\end{center}
\end{figure}

\begin{figure}[htp!]
\begin{center}
        \includegraphics[width=0.7\textwidth,width=0.7\columnwidth]{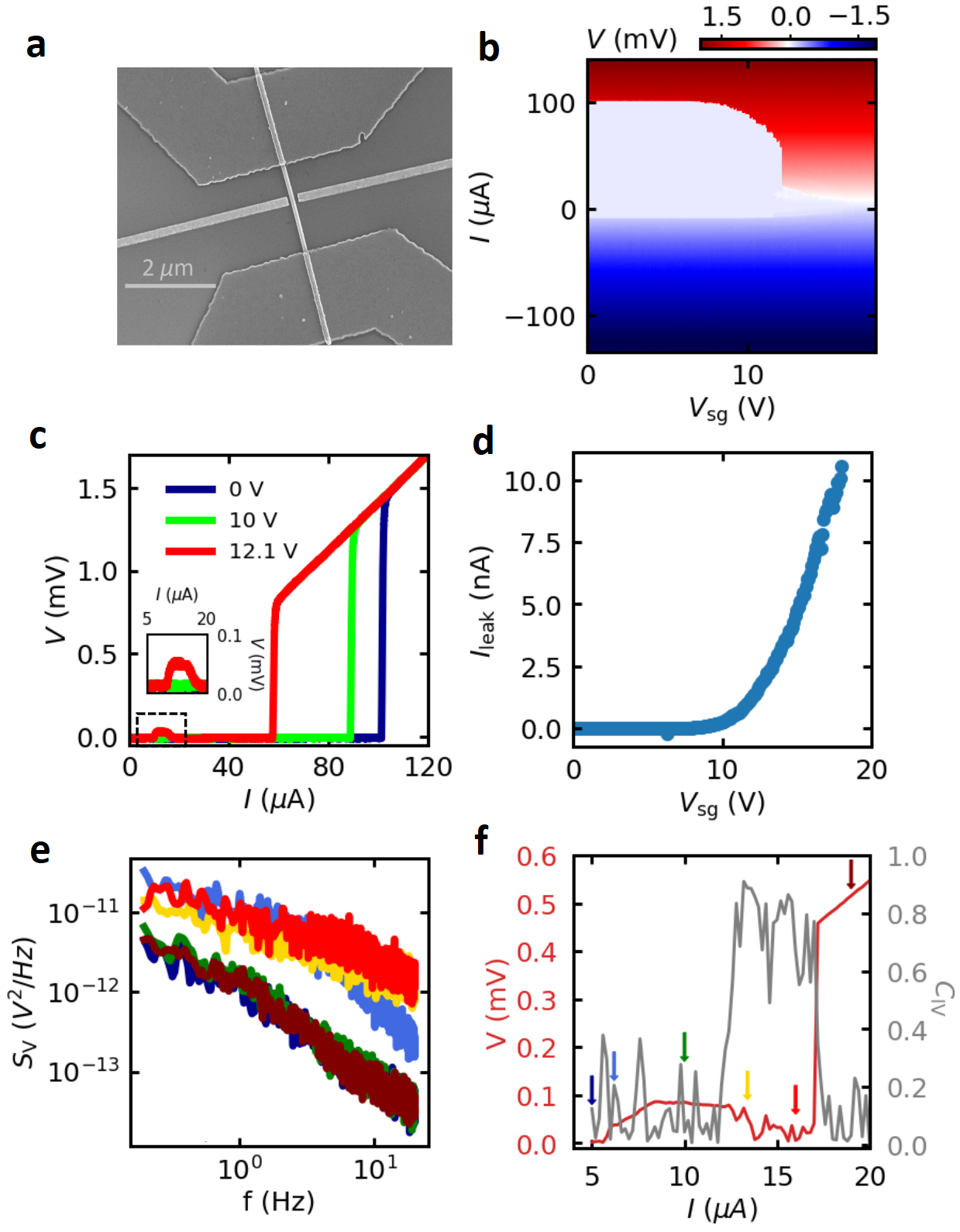}
	\caption{\label{fig:device_4} Investigation of the GCS and the noise correlations in device D ($L_{\mathrm{NW}}$=2\,$\mu$m). (a) SEM image of the investigated device. (b) $I-V$ characteristics of the NW device as a function of $V_{\mathrm{sg}}$. (c) Some selected $I-V$ curves are extracted from (b) at different gate voltages in the positive gate polarity. The $I-V$ curve measured at $V_{\mathrm{sg}}=12.1$\,V (red curve) shows the presence of the voltage state (surrounded by the black rectangle) below the switching current. The inset shows a magnification for the region in the $I-V$ curve surrounded by the black rectangle. (d) The corresponding $I_{\mathrm{leak}}$ as a function of $V_{\mathrm{sg}}$. (e) The NW noise spectra measured at $V_{\mathrm{sg}}=12.1$\,V at different bias current values indicated by the arrows in panel (f) that match the same color. (f) The $I-V$ curve that measured simultaneously with noise measurements and the corresponding coherence (at 3.35\,Hz) between the leakage current noise and the NW noise.}
\end{center}
\end{figure}

In the main text, we demonstrated the strong correlation between the leakage current and the fluctuations induced in the superconducting NW in a single device (device A). In this section, we demonstrate this correlation in three other devices (devices B, C, and D), as shown in SFigs.\,\ref{fig:device_2}, \ref{fig:device_3}, and \ref{fig:device_4}. The SEM images of the devices (panels a) show that they have different lengths of the NW segment (2$-$6.2$\mu$m), different interface areas with the Al contacts, and different shapes of the contacts. Panels b show that the GCS is realized for all devices with a corresponding enhancement in the leakage current (panels d). Interestingly, we noticed the presence of a finite voltage state below $I_{\mathrm{SW}}$ in all devices (panels c) at high gate voltages but with different shapes. In the case of devices C and D, the voltage state resembles the dissipative voltage state observed in superconducting NWs in a non-equilibrium state \cite{chen2014dissipative}.  Moreover, at the bias current values where the voltage fluctuation is present, a large enhancement in the level of the noise signal across the NW device (e panels) with a corresponding large coherence with leakage current noise (f panels) is observed.

\subsection{Time trace at $V_{\mathrm{sg}}$ = 5.35\,V at high bias current} 
\label{sssec:Time trace}
\begin{figure}[htp]
\begin{center}
        \includegraphics[width=0.55\textwidth,width=0.55\columnwidth]{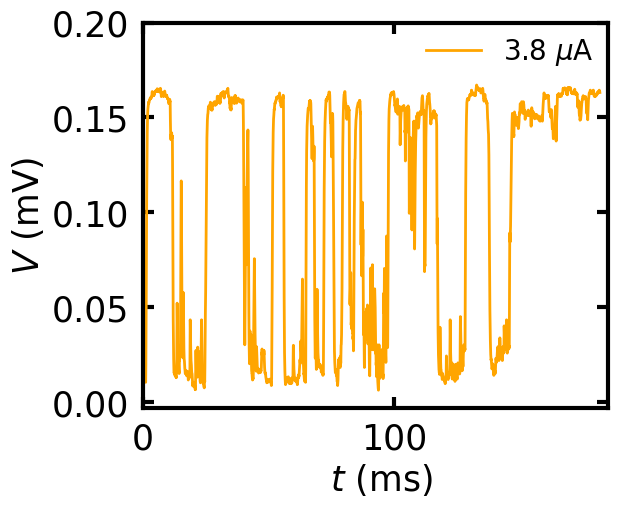}
	\caption{\label{fig:time_trace} Time trace of the voltage fluctuations at $I=3.8\,\mu$A and $V_{\mathrm{sg}}=5.35$\,V }
\end{center}
\end{figure}

As we discussed in the main text, the NW device at $V_{\mathrm{sg}}=5.35$\,V, fluctuates between the superconducting state and the intermediate voltage state in the bias current interval [2.4, 3.28]\,$\mu$A, while it fluctuates between the superconducting state and the normal state as the bias current increased beyond this interval as shown in the time trace in SFig.\,\ref{fig:time_trace} measured at $I=3.8\,\mu$A and $V_{\mathrm{sg}}=5.35$\,V.

\subsection{Gate dependence of the device in the first cooldown} 
\label{sssec:Gate dependence of the device in the first cooldown}
\begin{figure}[htp]
\begin{center}
\includegraphics[width=0.6\textwidth,width=0.6\columnwidth]{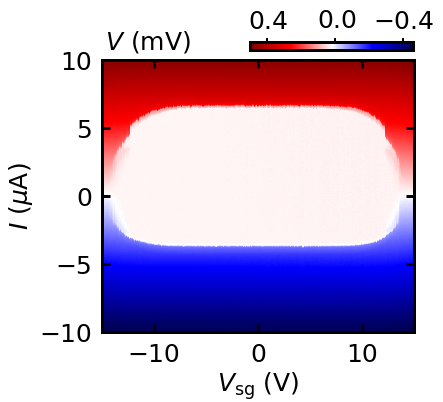}
	\caption{\label{fig:gate dependence_first cooldown} $I-V$ characteristics of the NW device as a function of $V_{\mathrm{sg}}$ in the first cool down.}
\end{center}
\end{figure}

In Fig.\,1b in the main text, it was demonstrated that the NW switching current became fully suppressed at $V_{\mathrm{sg}}=6$\,V. This measurement was carried out during a second cooldown. SFig.\,\ref{fig:gate dependence_first cooldown} shows the gate dependence of the device in the first cooldown in which $I_{\mathrm{SW}}$ was fully suppressed at $\simeq$ $\pm$15\,V.

\bibliography{references}